\documentclass[preprint,amsmath,amssymb,longbibliography,superscriptaddress,aps]{revtex4-2}
\usepackage{graphicx,bm}
\usepackage{appendix}
\usepackage{color}
\usepackage{mathtools}
\usepackage{soul}
\usepackage{float}
\usepackage{braket}

\newcommand{\rd}{\dot{\rho}}

\begin{document}

\title{Long-lived qubit entanglement by surface plasmon polaritons in a Weyl semimetal}

\author{M~Shoufie Ukhtary} 
\email{m.shoufie.ukhtary@brin.go.id}
\affiliation{Research Center for Quantum Physics, National Research and Innovation Agency (BRIN), South Tangerang 15314, Indonesia}

\author{Eddwi H. Hasdeo}
\affiliation{Research Center for Quantum Physics, National Research and Innovation Agency (BRIN), South Tangerang 15314, Indonesia}
\affiliation{Department of Physics and Materials Science, University of Luxembourg, L-1511 Luxembourg, Luxembourg}

\author{Andriyan B. Suksmono}
\affiliation{School of Electrical Engineering and Informatics, Institut Teknologi Bandung, Bandung 40132, Indonesia}
\affiliation{Research Collaboration Center for Quantum Technology 2.0, Bandung 40132, Indonesia}

\author{Ahmad R. T. Nugraha}
\affiliation{Research Center for Quantum Physics, National Research and Innovation Agency (BRIN), South Tangerang 15314, Indonesia}
\affiliation{Research Collaboration Center for Quantum Technology 2.0, Bandung 40132, Indonesia}

\begin{abstract}
We investigate spontaneous entanglement of two qubits mediated by nonreciprocal surface plasmon polaritons (SPPs) in a Weyl semimetal. In the absence of external magnetic fields, the topology of the Weyl semimetal even gives rise to nonreciprocal SPPs that are topologically protected and reside inside the photonic gap. We utilize this nonreciprocal SPP as a mediator of entanglement of two spatially-separated qubits. Our two main findings are: (1) the nonreciprocal SPP gives better quantum entanglement than the reciprocal one, and (2) the achieved entanglement is sufficiently long-lived compared to the entanglement using SPPs in conventional metals.
\end{abstract}

\date{\today}
\maketitle
\section{Introduction}
\label{sec:int}

Entanglement, which corresponds to the inseparability of quantum states between some quantum objects, is one of the fundamental concepts in quantum computing and quantum communication (or quantum key distribution) technologies. In the entangled state, the quantum state of one object influences the others, regardless of the distance between them. Using entanglement, the number of information embedded in the quantum system increases, thus improving the computation speed. Furthermore, long-distance entanglement is of great interest for developing efficient quantum circuits that require the information to transmit between spatially separated components~\cite{gonzalez2011entanglement,issah2021qubit,scully1999quantum,nielson2000quantum,o2007optical}. 

Recently, researchers have realized the long-distance entanglement between qubits by employing photons and plasmons as the mediator of the entanglement~\cite{gonzalez2011entanglement,otten2015entanglement,gangaraj2017robust}. In particular, when the qubits are coupled \emph{chirally} with an environment, the entanglement becomes more robust than the case of \emph{achiral} coupling since the coupling of qubits with the environmental degrees of freedom can become reduced. In the case of chiral coupling, the coupling strengths of a qubit with the forward and backward modes of the environment are different~\cite{gonzalez2015chiral}. The chiral coupling with the environment thus also provides less decoherence of the system~\cite{gangaraj2017robust,gonzalez2015chiral,pichler2015quantum}. A perfect chiral coupling may emerge in the nonreciprocal environment, where the qubits only couple to one degree of freedom of the environment~\cite{gangaraj2017robust}. The nonreciprocal environment has been realized in magnetized metals, where the surface plasmon polaritons (SPPs) propagate only in one direction perpendicular to the direction of bias voltage~\cite{gangaraj2017robust,gangaraj17-berry, silveirinha15-chern,monticone2020truly}. In particular, if the qubits are arranged at the interface between a magnetized metal and an opaque medium, the qubits only couple to the nonreciprocal SPPs, thus increasing the degree of entanglement~\cite{gangaraj2017robust}. Furthermore, researchers recently have shown that the chiral coupling by a nonreciprocal photonic environment increases the efficiency in energy transport between quantum emitters~\cite{gangaraj2022enhancement,doyeux2017giant}. The increase in efficiency originated from the larger cooperative decay rate between emitters than that of a reciprocal environment that increases the coupling between emitters. However, an external magnetic field must be applied to obtain a nonreciprocal environment. Such a requirement, in practice, is inconvenient for quantum circuitry. 

In this work, to avoid the requirement of external magnetic fields, we propose a Weyl semimetal as the intrinsic nonreciprocal environment. The Weyl semimetal is a three-dimensional metal having separated Dirac cones due to the broken symmetry~\cite{armitage2018weyl,vazifeh2013electromagnetic,zyuzin2012weyl}. In the case of broken time reversal symmetry, the Dirac cones are separated in wavevector as shown in Fig. \ref{fig:spp} (a)~\cite{armitage2018weyl,vazifeh2013electromagnetic,zyuzin2012weyl,hofmann16-surface}. Due to the topology of the Weyl semimetal, Hall current appears even without applied magnetic field~\cite{hofmann16-surface,burkov2014anomalous}.  The intrinsic Hall conductivity makes the Weyl semimetal optically anisotropic medium that supports nonreciprocal SPP, similar to the magnetized metal~\cite{chen2019optical,ukhtary2017negative,kotov2018giant,abdol2019tunable,bugaiko2020surface}. Some examples of the Weyl semimetal include pyrochlore ($\mathrm{Eu_2Ir_2O_7}$)~\cite{sushkov2015optical}, TaAs~\cite{lv2015experimental}, TaP~\cite{PhysRevB.101.014308}, $\mathrm{EuCd_2As_2}$~\cite{soh2019ideal} and NbAs~\cite{xu2015discovery}. It is noted that the bulk plasmon and SPP frequencies of the Weyl semimetal lie within the terahertz (THz) region in contrast to those of conventional metal, which mostly lie in the visible range~\cite{bugaiko2020surface,hofmann16-surface,maier2007plasmonics}. Thus, the use of SPP in the Weyl semimetal opens up the possibility to control the dynamics of qubits within THz frequency.

The rest of this paper is organized as follows.  In Sec.~\ref{sec:bulkandSPPdisp}, we start by investigating the bulk photonic dispersion inside the Weyl semimetal and obtaining the dispersion of SPP. We consider the interface between the normal metal and the Weyl semimetal. We found that there exists nonreciprocal SPP inside the common photonic band gap. Then, in Sec. III, we consider the open system consisting of two qubits, SPP, and the interaction between them. To understand the dynamics of the qubits, we derive the master equation within the Markov approximation. Finally, the entanglement is measured by so-called concurrence that is derived from the master equation. In Sec. IV, we show that the entanglement is more robust in the case of nonreciprocal SPP compared to that of the reciprocal one. Furthermore, since we work in the THz frequency, the decoherence rate is lower than that of the higher frequency, which gives a longer lifetime of the entanglement. Summary and perspectives are given in Sec. V.

\section{The bulk photonic dispersion and SPP dispersion}
\label{sec:bulkandSPPdisp}

\begin{figure}[t]
\begin{center}
\includegraphics[width=85mm]{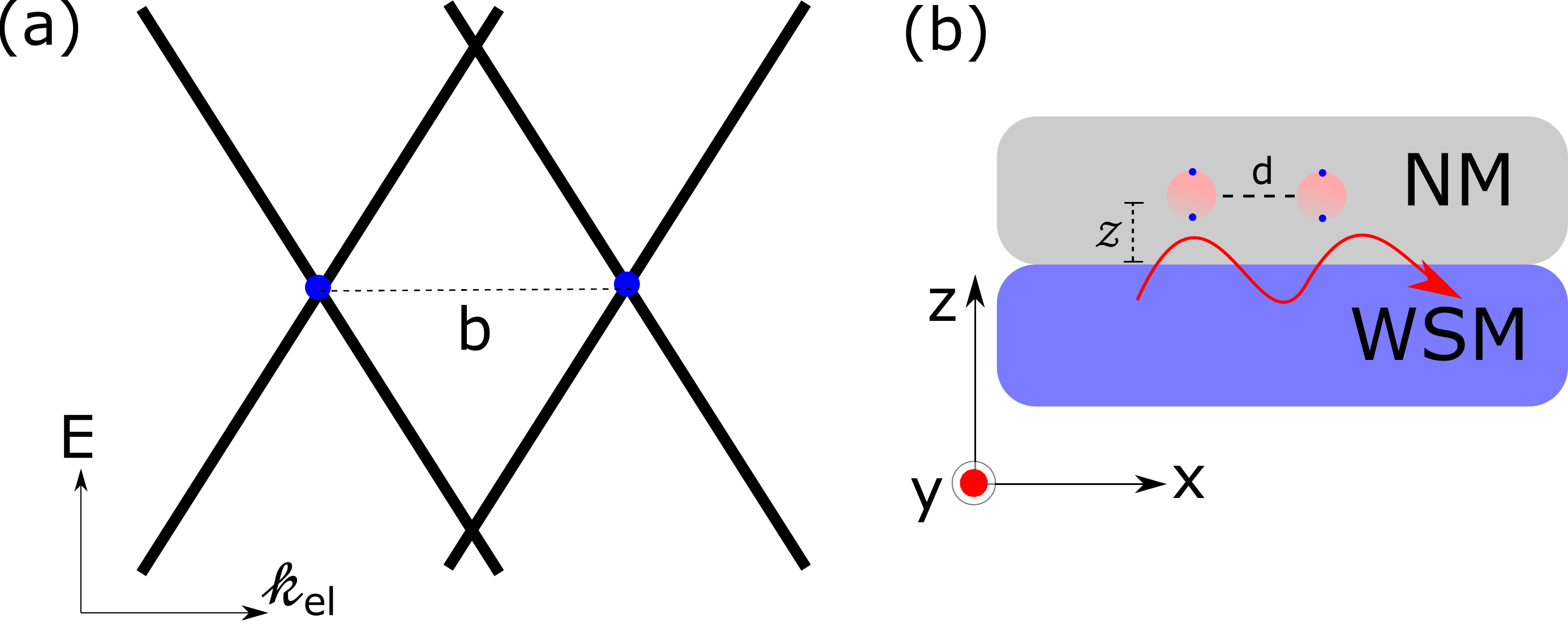}
\caption{(a) The electronic energy dispersion of the Weyl semimetal. The Weyl nodes are separated by the wavevector $\mathbf{b}$.  (b) Illustration of SPP propagating on the interface between the Weyl semimetal (WSM) and a normal metal (NM). The electromagnetic field of an SPP decays in the $z$ direction and propagates in the $x$ direction.  We also illustrate the qubits separated by a distance $d$. The qubits are placed above the WSM with a distance $\mathcal{Z}$.}\label{fig:spp}
\end{center}
\end{figure}

Let us begin by considering the propagation of light inside a Weyl semimetal. The electric displacement vector of a Weyl semimetal is given as follows \cite{chen13-axion,hofmann16-surface,vazifeh2013electromagnetic}: 

\begin{align}
    \mathbf{D} = \varepsilon_0 \varepsilon_\infty\left(1-\frac{\omega_p^2}{\omega^2 }\right)\mathbf{E}+\frac{ie^2}{4\pi^2\hbar\omega}(\nabla\theta)\times\mathbf{E},\label{eq:D}
\end{align}
where $\varepsilon_\infty$ is the background dielectric constant, $\omega_p$ is the bulk plasmon frequency of the Weyl semimetal and $\theta$ is called axion angle, which is determined by the separation of the Weyl nodes in the momentum space of electron. The coefficient in the first term of Eq. (\ref{eq:D}) is the conventional Drude permittivity, while the second one comes from the intrinsic Hall conductivity of the Weyl semimetal that is given in terms of axion angle~\cite{chen13-axion,vazifeh2013electromagnetic}. For the case of broken time reversal symmetry, the axion angle is defined as $\theta=2\mathbf{b}\cdot\mathbf{r}$ with $\mathbf{b}$ is the wave vector separating the Weyl nodes. It is noted that in the case of $\mathbf{b}=0$, we obtain the Dirac semimetal. Let us take the separation in the $y$ direction, thus we have $\theta = 2by$. Therefore, Eq. (\ref{eq:D}) can be cast compactly as the matrix equation as follows:
\begin{align}
\begin{bmatrix}
D_x\\
D_y\\
D_z
\end{bmatrix}=\varepsilon_0\begin{bmatrix}
\varepsilon_1 & 0 & i\varepsilon_2\\
0&\varepsilon_1 &  0\\
-i\varepsilon_2 & 0 & \varepsilon_1
\end{bmatrix}
\begin{bmatrix}
E_x\\
E_y\\
E_z
\end{bmatrix},
\label{eq:D3}
\end{align}
from which the dielectric tensor can be read as,
\begin{align}
\bar{\varepsilon}=\varepsilon_0\begin{bmatrix}
\varepsilon_1 & 0 & i\varepsilon_2\\
0&\varepsilon_1 &  0\\
-i\varepsilon_2 & 0 & \varepsilon_1
\end{bmatrix}
,
\label{eq:eps}
\end{align}
where
\begin{align}
\varepsilon_1 (\Omega)&= \varepsilon_\infty \left(1-\frac{1}
{\Omega^2}\right)\\
\varepsilon_2 (\Omega)&= \varepsilon_\infty\left(\frac{\Omega_b}{\Omega}\right)\label{eq:eps2}.
\end{align}
Here, we express $\varepsilon_1$ and $\varepsilon_2$ in terms of dimensionless quantities,
\begin{align}
    \Omega\equiv\frac{\omega}{\omega_p},~~~~~~~~~\Omega_b\equiv \frac{e^2b}{2\pi^2\varepsilon_0\hbar\omega_p\varepsilon_\infty}.
\end{align}
For simplicity, we simply write $\varepsilon_i(\Omega)$ as $\varepsilon_i$. 

The bulk photonic band is obtained by solving the Maxwell equations. In the matrix form, the source-free Faraday and Ampere laws are expressed as follows~\cite{silveirinha15-chern,gangaraj17-berry}:
\begin{align}
\begin{bmatrix}
0 & -\mathbf{k}\times\\
\mathbf{k}\times & 0
\end{bmatrix} \begin{bmatrix}
\mathbf{E}\\
\mathbf{H}
\end{bmatrix} =&~\omega\mathrm{M(\omega)}\begin{bmatrix}
\mathbf{E}\\
\mathbf{H}
\end{bmatrix}\label{eq:af},
\end{align}
where $\textrm{M}(\omega)$ is defined as follows:
\begin{align}
\mathrm{M(\omega)} =&~\begin{bmatrix}
\bar{\varepsilon}(\omega) & 0\\
0 & \mu_0{{\bar I}}
\end{bmatrix}\label{eq:M}.
\end{align}
Here, $\bar I$ is the $3\times3$ identity matrix and $\mathbf{k}=(k_x,0,k_z)$ is the wavevector of photon. Here, we take the photon propagation  in the $xz$ plane, thus $k_y=0$. We define $k_z=k\cos \phi$ and $k_x=k\sin\phi$ where $k$ is the magnitude of the photon wavevector and the angle $\phi$ corresponds to the angle of propagation that is measured from the $z$ axis. 

\begin{figure}[t]
\begin{center}
\includegraphics[width=75mm]{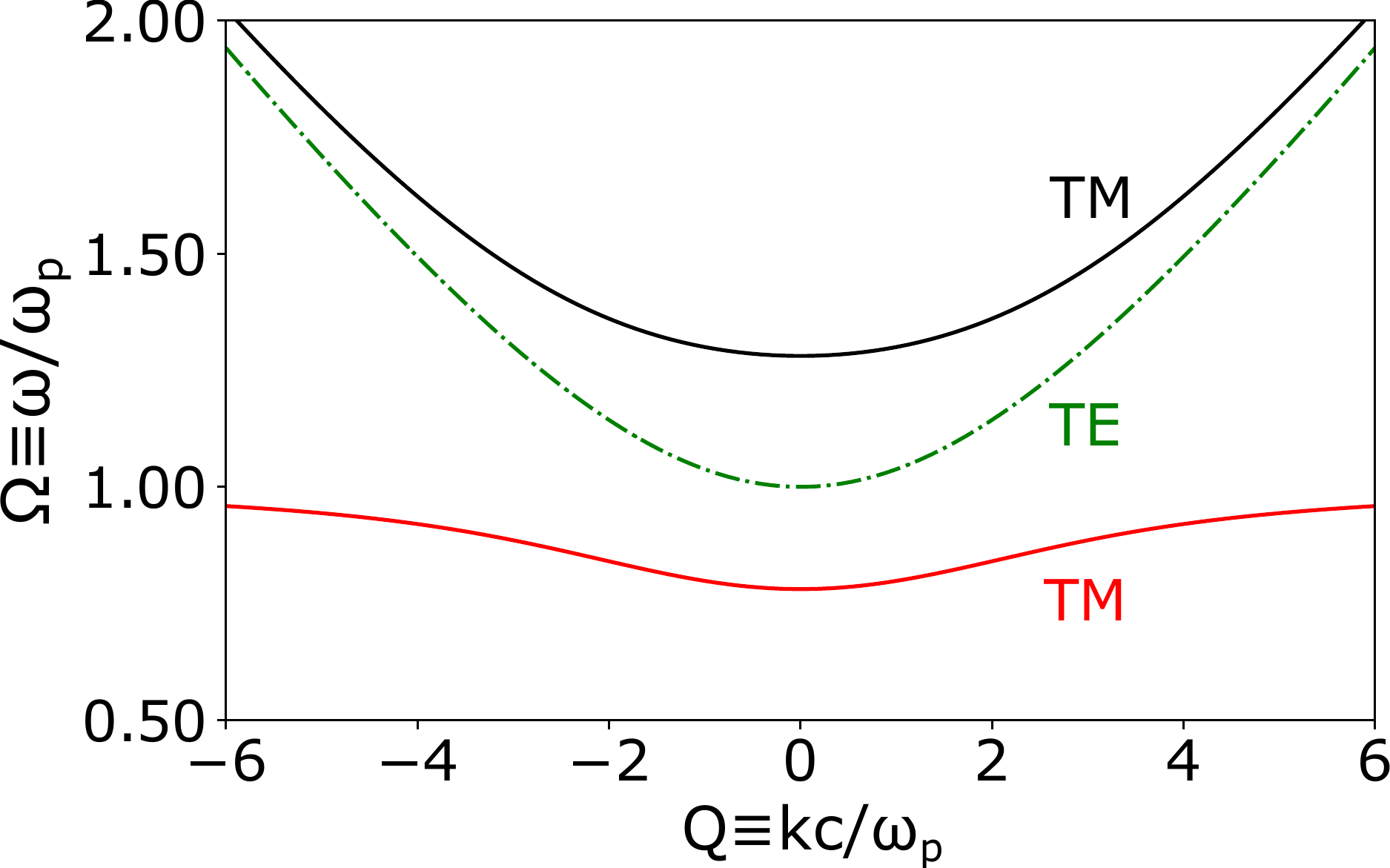}
\caption{The frequency of bulk photon propagating inside the Weyl semimetal with $xz$ plane of propagation. Here we use $\Omega_b=0.5$. }\label{fig:dispbl}
\end{center}
\end{figure}

We solve numerically the secular equation in Eq. (\ref{eq:af}) for $\omega$ as a function of photon wavevector $k$. In Fig. \ref{fig:dispbl}, we show $\Omega$ as a function of $Q\equiv kc/\omega_p$ of a photon propagating inside the Weyl semimetal with the wave vector in the $xz$ plane. There are three distinct branches. The lowest and the highest branches correspond to the transverse magnetic (TM) waves that are separated by a band gap, while the middle branch corresponds to the transverse electric (TE) wave. The wave function of the TM wave is given by,
\begin{align}
\mathbf{f_k}^\textrm{TM}= \begin{bmatrix}
\mathbf{E}^\textrm{TM}\\
\mathbf{H}^\textrm{TM}\end{bmatrix}= \begin{bmatrix}
-\displaystyle\frac{H_0}{\omega}\left(\bar{\varepsilon}\right)^{-1}(\mathbf{k}\times\hat{\mathbf{y}})\\
H_0\hat{\mathbf{y}}\end{bmatrix},\label{eq:TM}
\end{align}
while that for TE wave,
\begin{align}
\mathbf{f_k}^\textrm{TE}= \begin{bmatrix}
\mathbf{E}^\textrm{TE}\\
\mathbf{H}^\textrm{TE}\end{bmatrix}= \begin{bmatrix}
E_0\hat{\mathbf{y}}\\
\displaystyle\frac{E_0}{\omega\mu_0}(\mathbf{k}\times\hat{\mathbf{y}})
\end{bmatrix}.\label{eq:TE}
\end{align}
It is noted that in the case of TE wave, $\Omega$ is determined solely by $\varepsilon_1$. We check it easily by looking at the Eq. (\ref{eq:D3}) and $\mathbf{f_k}^\textrm{TE}$. Since the TE wave has only an electric field component in the $y$ direction, thus the displacement vector is only given by $D_y=\varepsilon_1(\omega)E_y$, which depends only on the Drude permittivity of the Weyl semimetal, $\varepsilon_1(\omega)$. Thus, $\Omega$ is simply given by $\Omega^{\textrm{TE}}=\sqrt{1+Q^2/\varepsilon_\infty}$.

Similar to the electronic system, the photonic band structure  might possess a topological phase, which is characterized by the Chern number. Silveirinha has formulated the definition of the Chern number for continuous media that is used to classify the topology of the media~\cite{silveirinha15-chern}. In particular, he showed that the interface between trivial and nontrivial media hosts surface states of photon or SPP that is topologically protected~\cite{silveirinha15-chern,gangaraj17-berry}. The dispersion of the SPP lies inside the common band gap, thus it would not decay into radiation to the surrounding media. More importantly, he pointed out that the bulk-edge correspondence also applies in the continuous media, that is the difference of the Chern number between the two media corresponds to the number of surface state~\cite{silveirinha15-chern,silveirinha2016bulk}. 

\begin{figure}[t]
\begin{center}
\includegraphics[width=85mm]{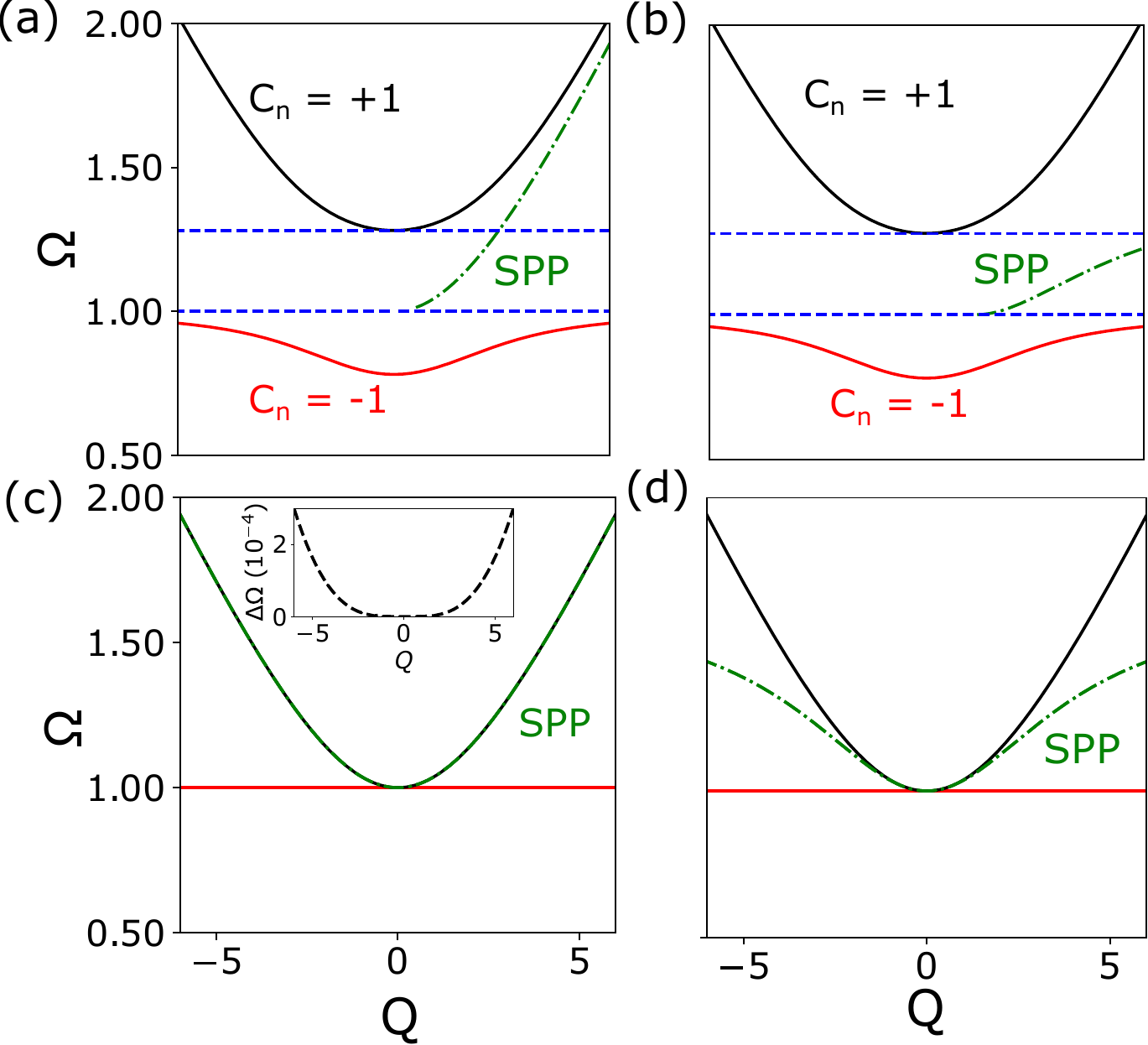}
\caption{ (a) The frequency of TM bulk photon propagating inside the Weyl semimetal and the frequency of SPP at the interface between the Weyl semimetal and normal metal. Here the bulk plasmon frequency of the normal metal (silver) is $293\omega_p~(9.6$ eV). (b) The same with (a), but for  the bulk plasmon frequency of the normal metal of 5$\omega_p$. (c) and (d) The same with (a) and (b), respectively, but we use the Dirac semimetal ($\Omega_b=0$, instead of the Weyl semimetal ($\Omega_b\neq 0$)). The inset in (c) is the frequency difference between the frequency of the SPP and the bulk TM photon. The order of $\Delta\Omega$ is $10^{-4}$ and $\Delta\Omega$ increases with increasing $Q$.  Here, we use $\Omega_b=0.5$. }\label{fig:disp}
\end{center}
\end{figure}

It was shown that the photon Chern number  inside a metal is not trivial when a magnetic field is applied, due to the breaking of the time reversal symmetry \cite{silveirinha15-chern,gangaraj17-berry}. In this case, the dielectric tensor has off-diagonal terms due to the Hall conductivity. Let us show that the Weyl semimetal has a non-trivial topology that occurs due to the anomalous Hall conductivity. The Chern number is defined as follows~\cite{silveirinha15-chern}:
\begin{align}
   C_n&= \frac{1}{2\pi}\oint\limits_{k=\infty}\mathbf{A_k}\cdot \mathbf{dl} - \frac{1}{2\pi}\oint\limits_{k=0^+}\mathbf{A_k}\cdot \mathbf{dl},\label{eq:ch}\\
   &=\lim_{k\rightarrow \infty}(A_\phi k) - \lim_{k\rightarrow 0^+}(A_\phi k)\label{eq:cnum}
\end{align}
where  $\mathbf{A_k}$ is the photon Berry connection. The Berry connection is determined by the wave function given in Eqs. (\ref{eq:TM}) and (\ref{eq:TE}). The Berry connection is defined as \cite{silveirinha15-chern,gangaraj17-berry}:
\begin{align}
    \mathbf{A_k}=\frac{i{\mathbf{f_k^*}\cdot\partial_\omega(\omega\mathrm{M}(\omega))\nabla_k\mathbf{f_k}}}{{\mathbf{f_k^*}\cdot\partial_\omega(\omega\mathrm{M}(\omega))\mathbf{f_k}}},\label{eq:ak}
\end{align}
where $\mathrm{M}(\omega)$ is a matrix defined by  $\bar{\varepsilon}(\omega)$ and $\mu_0$ in Eq.~(\ref{eq:M}).
By introducing the matrix $\mathrm{M}(\omega)$, the normalization factor in Eq.~(\ref{eq:ak}) (denominator of Eq.~ (\ref{eq:ak})) corresponds to the energy density of electromagnetic wave.
When we use polar coordinate, $A_\phi$ is given as follows:
\begin{align}
    {A_\phi}&=\frac{\mathrm{Re}(i\mathbf{E^*}\cdot\partial_\omega(\omega\bar{\varepsilon}(\omega))\frac{1}{k}\partial_\phi \mathbf{E} + i\mathbf{H^*}\cdot\mu_0\frac{1}{k}\partial_\phi \mathbf{H})}{\mathbf{E^*}\cdot\partial_\omega(\omega\bar{\varepsilon}(\omega))\mathbf{E} + \mathbf{H^*}\cdot\mu_0 \mathbf{H}}\label{eq:ath}.
\end{align}
In going from Eq. (\ref{eq:ch}) to (\ref{eq:cnum}), we use that $\mathbf{A_k}$ does not depend on $\phi$ and $\mathbf{A_k}$ in the radial direction vanishes. In Figs. \ref{fig:disp}(b) and (c), we show the band structure for the TM wave together with the calculated Chern number. The Chern number $+(-)~ 1$ is obtained for the upper (lower) energy TM branch. Then non-trivial Chern number for the branch below the band gap gives us a clue about the possible existence of SPP when the Weyl semimetal is attached to a trivial material. This shows that the Weyl semimetal is intrinsic photonic topological material since we do not need to apply a magnetic field for the non-trivial Chern number. It is noted that the total Chern number of the two branches is always zero. Similarly, in the case of normal metal without $\varepsilon_2$ and of TE wave, we only have one branch with a trivial Chern number. It is noted that we do not consider the non-locality of the dielectric function when we calculate the Chern number as is usually performed to obtain an integer Chern number for the lower branch~\cite{silveirinha15-chern,gangaraj17-berry,pakniyat2022chern}. However, we obtain integer Chern numbers for both branches, which we prove analytically in Appendix B. From the calculation, we infer that the functional form of $\varepsilon_2$ of the WSM, which is simply given by $1/\omega$~[Eq.~(5)], allows the integer Chern number even without considering non-locality. 
 
To exploit this topological SPP, we attach the Weyl semimetal to a normal metal as shown in Fig. \ref{fig:spp} (b). The normal metal has a larger bulk plasmon frequency than that of the Weyl semimetal. This structure gives $\Delta C_n=1$, which means that there is a band of unidirectional SPP lying inside the common band gap, which is similar to the case of magnetized metal. To derive the dispersion of SPP, we define the electromagnetic field of the SPP in both media. Inside the normal metal ($z>0$), we have the following fields:
\begin{align}
    H^{(1)}_y (x,z)&= H^{(1)}_0 e^{iqx}e^{-\kappa_1 z}\label{eq:hy2}\\
    E^{(1)}_x (x,z)&= \frac{i\kappa_1 H^{(1)}_0}{\omega\varepsilon_0\varepsilon_{m}}e^{iqx}e^{-\kappa_1 z}\label{eq:ex2}\\
    E^{(1)}_z (x,z)&= -\frac{qH^{(1)}_0}{\omega\varepsilon_0\varepsilon_{m}}e^{iqx}e^{-\kappa_1 z},\label{eq:ez2}
\end{align}
where $q$ is the wave vector of SPP and $\varepsilon_{m}$ is the Drude permittivity for the normal metal. Inside the WSM ($z<0$), the fields are given as follows:
\begin{align}
    H_y^{(2)}(x,z) &= H_0^{(2)}e^{iqx}e^{\kappa_2 z}\label{eq:hy3}\\
    E_x^{(2)}(x,z) &= \frac{H_0^{(2)}}{\omega\varepsilon_0}\frac{1}{\varepsilon_1^2-\varepsilon_2^2}\left(-i\kappa_2\varepsilon_1+ i\varepsilon_2 q\right)e^{iqx}e^{\kappa_2 z}\label{eq:ex3}\\
    E_z^{(2)}(x,z) &= \frac{H_0^{(2)}}{\omega\varepsilon_0}\frac{1}{\varepsilon_1^2-\varepsilon_2^2}\left(\kappa_2\varepsilon_2  - \varepsilon_1 q\right)e^{iqx}e^{\kappa_2 z}\label{eq:ez3}.
\end{align}
Here, $\kappa_i$ is the decay constant of the fields in the $z$ direction. By substituting $E_x(x,z)$ and $E_z(x,z)$ to the electromagnetic wave equation
we obtain the expression for $\kappa_i$ as follows:
\begin{align}
    \kappa_1&=\sqrt{q^2-(\omega/c)^2\varepsilon_{m}}\\
    \kappa_2&=\sqrt{q^2+(\omega/c)^2\varepsilon_2^2/\varepsilon_1-(\omega/c)^2\varepsilon_1}.
\end{align}
The dispersion of the SPP can be obtained by substituting Eqs. (\ref{eq:hy2}) - (\ref{eq:ez3}) to the boundary conditions which state that the tangential electric and magnetic fields are continuous at the boundary. Let us consider that $E_x^{(1)}(x,0)=E_x^{(2)}(x,0)$ and $H_y^{(1)}(x,0)=H_y^{(2)}(x,0)$. Thus, we have the following equation,
\begin{align}
    \frac{\kappa_1}{\varepsilon_{m}}=\frac{1}{\varepsilon_1^2-\varepsilon_2^2}\left(-\kappa_2\varepsilon_1+ \varepsilon_2q\right).
\end{align}
In Figs. (\ref{fig:disp}) (a) and (b), we show the dispersion of SPP (dash-dotted line) together with the bulk photonic bands for the TM wave. The common band gap is bordered by the two dashed lines. The frequency range of the common band gap $(\Omega_\textrm{gap})$ is given by,
\begin{align}
    1<\Omega_\textrm{gap}<1/2(\Omega_b+\sqrt{4+\Omega_b^2})\label{eq:gap}.
\end{align}
It is important to note that $\omega_p$ of the Weyl semimetal lies within the THz range. For example, in the case of $\mathrm{Eu_2Ir_2O_7}$, $\omega_p=66 $~THz, which corresponds to a frequency of 10 THz.
In (a), we use silver for the normal metal whose bulk plasmon frequency is $293\omega_p~(9.6$ eV), while in (b), we use a normal metal with a bulk plasmon frequency of $5\omega_p~$ that is closer to the  $\omega_p$. In both cases, we found a nonreciprocal SPP branch inside the common band gap that corresponds to $\Delta C_n=1$. Therefore, the frequency of the SPP lies also in a THz range, which can be used to entangle qubits whose transition frequency is in a THz range. It is noted that we look at the solutions having real $\kappa_i$, which means that the fields decay in both media.    

The dispersion of SPP tends to saturate near the bulk plasmon frequency of metal~\cite{maier2007plasmonics}. The dispersion of SPP in (a) is steeper than that in (b), since the bulk plasmon frequency of the normal metal is much higher in (a) than that in (b). Since the SPP originates from the contrast of the topological invariance between the two media, the SPP is robust against any disorders on the surface~\cite{monticone2020truly,gangaraj17-berry}. On the other hand,  SPP that appears in the interface between two trivial media, might not be robust against disorder. Let us take the interface between a normal metal and a Dirac semimetal by setting $\varepsilon_2=0$. The dielectric tensor of the Dirac semimetal is similar to that of a normal metal, thus it is a trivial medium. Nevertheless, we obtain the SPP that satisfies the boundary conditions. The dispersions of bulk photon and the SPP are given in Figs. \ref{fig:disp} (c) and (d). There is no common band gap (the band gap is defined only for the normal metal) and the dispersion of SPP appears above $\omega_p$ but lower than the frequency of bulk plasmon of the normal metal so that $\varepsilon_1$ has the opposite sign with $\varepsilon_m$. In particular, the SPP is reciprocal. In (c), the dispersion of SPP is close to the dispersion of the bulk photon, while in (d) it is close to the bulk photon at small $Q$. The dispersion is also steeper in (c) than that in (d), since the dispersion saturates at a much higher frequency in (c).

\section{The qubit dynamics and entanglement}
In this section, we will consider the coupling of a quantum emitter with the nonreciprocal SPP in the Weyl semimetal. The coupling is given by the spontaneous emission of the SPP by the quantum emitter. The quantum emitter is represented by a two-level system or qubit. Since we focus on the energy range within the common band gap, the quantum emitter couples only to the SPP, not to the bulk photon. The coupling of a qubit with the SPP enables the possibility to entangle two qubits separated by a distance, where the SPP acts as a mediator. This phenomenon has been studied in the case of plasmonic wave guide~\cite{gonzalez2011entanglement} and also in the case of magnetized metal~\cite{gangaraj2017robust}. 

Let us consider a system consisting of SPP  and two qubits. The qubits are placed on the top of the Weyl semimetal by a distance $\mathcal{Z}$ as shown in Fig. \ref{fig:spp} (b).  We set $\mathcal{Z}$ to be $1$ nm. To measure the entanglement between the qubits, we first calculate the occupation of the qubits. Let us write down the Hamiltonian as follows~\cite{gangaraj2017robust,gonzalez2013theory}:
\begin{align}
    H &= \sum\limits_{i=1}^2 \hbar\omega_0 \sigma_i^+\sigma^-_i+\sum\limits_k\hbar\omega_k(a_k^{\dagger}a_k +\frac{1}{2})+H_{I}\nonumber\\
    &=H_S +H_E + H_I,
\end{align}
where we consider that the system consists of two qubits with Hamiltonian $H_S$, the environment is the SPP with  Hamiltonian $H_E$ and the interaction between the qubit and SPP is given by Hamiltonian $H_I$. $\sigma^+_i$ and $\sigma^-_i$ are the raising  and lowering operators of each qubit, while $a_k^\dagger$ and $a_k$ are the creation and annihilation operators of the SPP with wave vector $k$. Let us now consider the interaction between a qubit with the SPP. The interaction Hamiltonian is derived as follows~\cite{archambault2010quantum,gangaraj2017robust,salasnich2017quantum,gonzalez2013theory}:
\begin{align}
    H_I = \int dr \Psi^\dagger (r,t)~\mathbf{d\cdot E} ~\Psi (r,t),\label{eq:hi}
\end{align}
where $\mathbf{d}$ is the dipole moment of the qubit and $\Psi(r,t)$ is the field operator of an electron in the two-level system.


The electric field of SPP is given by the quantized electric field as follows~\cite{archambault2010quantum,gonzalez2011entanglement}:
\begin{align}
    \mathbf{E(r,t)} = i\sum\limits_{k}\sqrt{\frac{\hbar\omega}{2\varepsilon_0 S}}\mathbf{u_k}(z)a_k e^{ikx-i\omega t} + \mathrm{h.c},\label{eq:eq}
\end{align}
where $S$ is the area of the quantization of SPP and $\mathbf{u_k}(z)$ represents the polarization of the SPP. Since the qubits are placed on the top of the Weyl semimetal, $\mathbf{u_k}(z)$ is expressed as follows~\cite{archambault2010quantum}:
\begin{align}
    \mathbf{u_k}(z)=\alpha e^{-\kappa_1z}\left(\mathbf{\hat{x}}-\frac{k}{ i\kappa_1}\mathbf{\hat{z}}\right)\label{eq:u}.
\end{align}
 By taking the rotating wave approximation, the total interaction Hamiltonian is expressed as follows:
\begin{align}
    H_I = \sum\limits_{i=1}^{2}\sum\limits_{k}\left(g_{ki}^* a_k^\dag\sigma_i^- e^{-i(\omega_0-\omega_k)t}+ g_{ki} a_k\sigma_i^+ e^{i(\omega_0-\omega_k)t}\right),\label{eq:hi2}
\end{align}
where the coupling constant $g_{ki}$ is defined as,
\begin{align}
    g_{ki}=-i\sqrt{\frac{\hbar\omega}{2\varepsilon_0 S}}\mathbf{u_k}(z)\cdot \mathbf{d_{12}}e^{ikr_i}.
\end{align}
Here, $r_i$ is the position of a qubit $i$ and $\mathbf{d_{12}}=\int dr \phi_1(r)\mathbf{d}\phi_2(r)$ is the dipole moment matrix element,  where $\phi_i(r)$ is the wave function of an electron in the energy level $i$. The $\alpha$ factor in Eq. (\ref{eq:u}) is defined so that the total energy of the SP is given in the form of oscillator harmonic energy~\cite{archambault2010quantum}. We have $\alpha^{-2}$ as
\begin{widetext}
\begin{align}
    (\alpha^2)^{-1}=\frac{1}{2\kappa_1}\left[\Tilde{\varepsilon}_m\left(1+\frac{k^2}{\kappa_1^2}\right)+\frac{\omega^2}{c^2}\left|\frac{\varepsilon_m}{\kappa_1}\right|^2\right]+\frac{1}{2\kappa_2}\left[\Tilde{\varepsilon}_1\left(1+\left|\frac{\kappa_2\varepsilon_2-k\varepsilon_1}{-\kappa_2\varepsilon_1+k\varepsilon_2}\right|^2\right)+\frac{\omega^2}{c^2}\left|\frac{\varepsilon_2^2+\varepsilon_1^2}{-\kappa_2\varepsilon_1+k\varepsilon_2}\right|^2\right],
\end{align}
\end{widetext}
where $\Tilde{\varepsilon}_i(\omega)$ is defined by,
\begin{align}
    \Tilde{\varepsilon}_i(\omega)=\frac{\partial}{\partial\omega}\omega\varepsilon_i(\omega).
\end{align}

The occupation of the qubits is governed by the master equation, which is the time evolution of the density operator of the system. For the spontaneous entanglement, the master equation~\cite{manzano2020short} of our system is given as follows:
\begin{widetext}
\begin{align}
    \partial_t\rho(t) =-\frac{i}{\hbar}\left[H_S,\rho(t)\right]-\frac{1}{\hbar^2}\left\{\sum\limits_{ij}\left(\Gamma_{ij}\left[\sigma_i^+,\sigma_j^-\rho(t)\right]+\Gamma_{ij}^{*}\left[\rho(t)\sigma_j^+,\sigma_i^-\right]\right)\right\}.
\end{align}
\end{widetext}

By defining the bases $ \ket{d}=\ket{e_1,g_2},\ket{u}=\ket{g_1,e_2}, \ket{g}=\ket{g_1,g_2},\ket{e}=\ket{e_1,e_2}$, where $g_i$ and $e_i$ correspond to the ground and excited states of qubit-i, and the following operations: $\sigma^{+}_i\ket{g_i}=\ket{e_i}$, $\sigma^{+}_i\ket{e_i}=0$, $\sigma^{-}_i\ket{e_i}=\ket{g_i}$ and $\sigma^{-}_i\ket{g_i}=0$, we obtain the following matrix elements for the density operator for the two qubits system,
\begin{align}
 \rd_{dd}&=-\frac{1}{\hbar^2}\left(2\textrm{Re}(\Gamma)\rho_{dd}+\Gamma_{12}\rho_{ud}+\Gamma_{12}\rho_{du}\right) \nonumber\\
 \rd_{uu}&=-\frac{1}{\hbar^2}\left(2\textrm{Re}(\Gamma)\rho_{uu}+\Gamma_{21}\rho_{du}+\Gamma_{21}^{*}\rho_{ud}\right) \nonumber\\
 \rd_{ud}&=-\frac{1}{\hbar^2}\left(2\textrm{Re}(\Gamma)\rho_{ud}+\Gamma_{21}\rho_{dd}+\Gamma_{12}^{*}\rho_{uu}\right) \nonumber\\
 \rd_{du}&=-\frac{1}{\hbar^2}\left(2\textrm{Re}(\Gamma)\rho_{du}+\Gamma_{12}\rho_{uu}+\Gamma_{21}^{*}\rho_{dd}\right) \nonumber\\
 \rd_{gg}&=-\frac{1}{\hbar^2}\Bigg(-2\textrm{Re}(\Gamma)(\rho_{dd}+\rho_{uu})-\Gamma_{12}\rho_{ud}\nonumber\\
 &-\Gamma_{21}\rho_{du}-\Gamma_{12}^{*}\rho_{du}-\Gamma_{21}^{*}\rho_{ud}\Bigg),\label{eq:rhod}
\end{align}
where $\Gamma=\Gamma_{ii}$. Here, we assume that the system is initially at the state $\ket{d}$. The $\Gamma_{ij}$ parameter given in units of $\mathrm{(eV)^2s}$ is the total coupling constant of a qubit with the environment and is expressed as follows:
\begin{align}
    \Gamma_{ij}&=\lim_{\eta\to 0}\sum\limits_k\int\limits_0^\infty ds e^{i(\omega_0-\omega_k)s-\eta s}g_{ki}g^*_{kj}\nonumber\\&=\sum\limits_k g_{ki}g^*_{kj} \left[\pi\delta(\omega_0-\omega_k)+
    \frac{i}{\omega_0-\omega_k}\right]\label{eq:gg}.
\end{align}
It is noted that the $\Gamma_{ij}$ parameter in Eq.~(\ref{eq:gg}) is derived quantum mechanically since we express the electric fields of the SPP in terms of field operators. However, the evaluation of the coupling constant can also be done classically by considering the pole contribution of the electromagnetic Green's tensor~\cite{gangaraj2017robust}. Archambault et al. have previously shown the equivalence of both methods in their paper~\cite{archambault2010quantum}. 

In the case of nonreciprocal SPP, we do not have left-going SPP. The $\Gamma_{ij}$ parameter corresponds to the excitation of qubit $i$ and relaxation of qubit $j$, meaning that a SPP is propagating from qubit $j$ to $i$. If we take $r_2>r_1$, then $\Gamma_{12}$ corresponds to the left-going SPP, while $\Gamma_{21}$ corresponds to the right-going SPP. Since we do not have left-going SPP, thus we can set $\Gamma_{12}=0$ in the differential equations (\ref{eq:rhod}). Similar to the results for the magnetized metal~\cite{gangaraj2017robust}, the solutions for the nonreciprocal case are
\begin{align}
    \rho_{dd}&=e^{-\displaystyle\frac{2\textrm{Re}(\Gamma)t}{\hbar^2}}\nonumber\\
     \rho_{uu}&=t^2\frac{\left|\Gamma_{21}\right|^2}{\hbar^4}e^{-\displaystyle\frac{2\textrm{Re}(\Gamma)t}{\hbar^2}}\nonumber\\
     \rho_{ud}&=-t\frac{\Gamma_{21}}{\hbar^2}e^{-\displaystyle\frac{2\textrm{Re}(\Gamma)t}{\hbar^2}}\nonumber\\
     \rho_{du}&=\rho_{ud}^*.\label{eq:solnres}
\end{align}

To measure the entanglement, we use the so-called concurrence function $\mathcal{C}$ introduced by Wootters~\cite{wootters2001entanglement}. The $\mathcal{C}$ function has a value ranging from 0 for the unentangled state to 1 for the completely-entangled state. In our system, the $\mathcal{C}$ function is expressed as:
\begin{equation}
    \mathcal{C}= 2|\rho_{du}|=2|\rho_{ud}|.
    \label{eq:concur}
\end{equation}

\section{Qubit Entanglement by the nonreciprocal SPP}

To calculate $\mathcal{C}$, let us determine $\Gamma_{ij}$ for our system. Since the SPP propagates unidirectionally in the $x$ direction, we set $k_y = 0$ and only the summation in $k_x\equiv q$ is considered as follows:
\begin{align}
    \Gamma_{ij}&=\sum\limits_{q} g_{qi}g^*_{qj} \left[\pi\delta(\omega_0-\omega_q)+\frac{i}{\omega_0-\omega_q}\right]\label{eq:gij1}\\
    &=\frac{L_x}{2\pi}\int dq g_{qi}g^*_{qj} \left[\pi\delta(\omega_0-\omega_q)+\frac{i}{\omega_0-\omega_q}\right]\\
    &=\frac{\hbar}{4\pi\varepsilon_0L_y}\int~dq~ \omega_q |\mathbf{u_q}(z)\cdot \mathbf{d_{12}}|^2e^{-iq(r_j-r_i)}\nonumber\\&\times\left[\left(\frac{\partial\omega_q}{\partial q}\right)_{q=q_0}^{-1}\pi\delta(q_0-q)+\frac{i}{\omega_0-\omega_q}\right]
    \label{eq:gamin},
    \end{align}
where $q_0$ is the wave vector of SPP for $\omega_q=\omega_0$. Here, $L_y$ is the sample width, which we set to be $L_y=1~\mu$m. The qubits are separated by a distance of $d=2\pi /q_0$. By using the contour method for evaluating the second integral in Eq.~(\ref{eq:gamin}) with contour given by an infinitesimal half-circle around $\omega_0$, we found that the result of the first and second integrals in Eq.~(\ref{eq:gamin}) are the same so that $\Gamma_{ij}$ is given as follows:
\begin{align}
    \Gamma_{ij}=\frac{\hbar}{2
    \varepsilon_0L_y}F(\omega_0,q_0),\label{eq:gij}
\end{align}
where
\begin{align}
    F(\omega_q,q)=\omega_q |\mathbf{u_q}(z)\cdot \mathbf{d_{12}}|^2e^{-iq(r_j-r_i)}\left(\frac{\partial\omega_q}{\partial q}\right)^{-1}\label{eq:f}.
\end{align}
Substituting Eq.~(\ref{eq:gij}) into Eq.~(\ref{eq:solnres}), we can evaluate $\mathcal{C}$ [Eq.~\eqref{eq:concur}] as a function of time. It is noted that $\Gamma\equiv \Gamma_{ii}$ is real-valued as given by Eq.~(\ref{eq:f}). We consider an ideal SPP, in which the damping of SPP is neglected. Thus, the distance between two qubits appears as a phase in Eq.~(\ref{eq:f}), which does not affect $\mathcal{C}$.
\begin{figure}[t]
\begin{center}
\includegraphics[width=85mm]{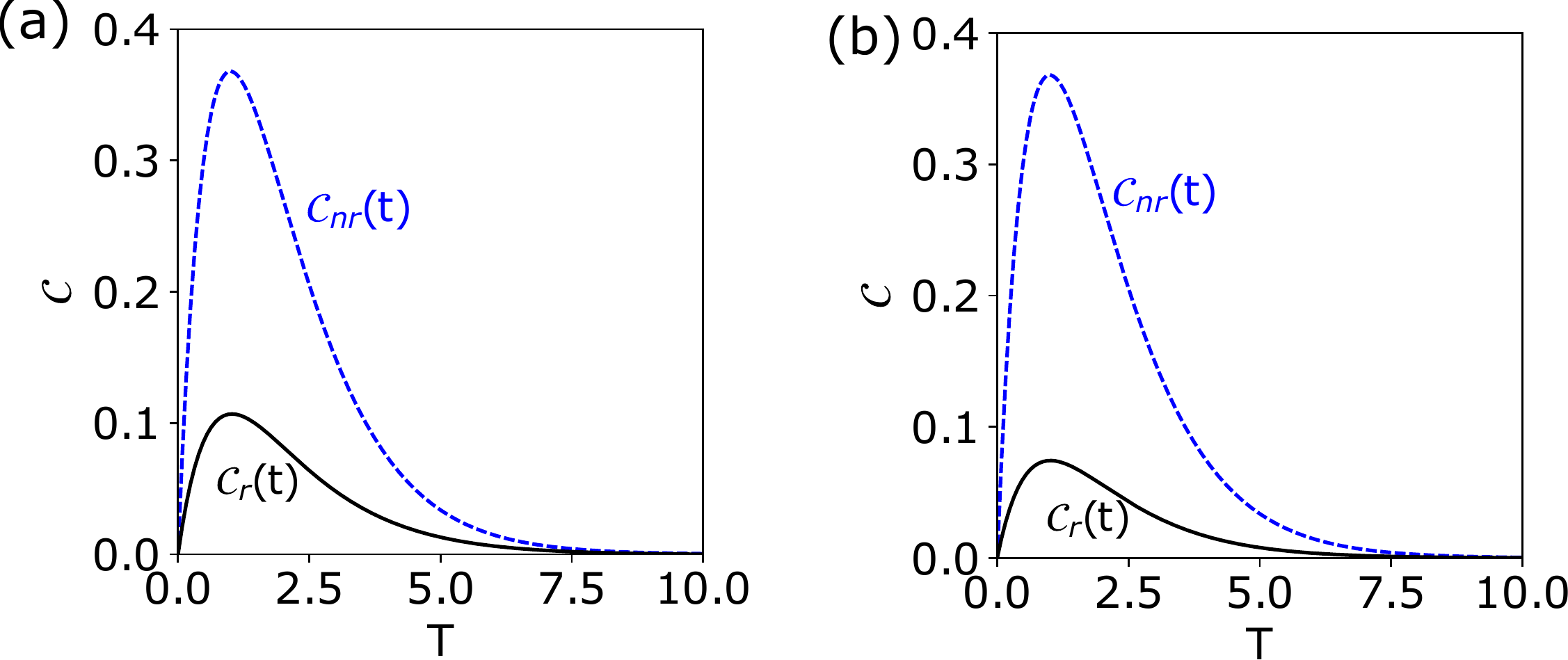}
\caption{The concurrence for the Weyl semimetal (dashed line) and the Dirac semimetal (solid line) as a function of normalized time $T=2\Gamma t/\hbar^2$, where $\Gamma$ is defined for each material. (a) The bulk plasmon frequency of the normal metal is $5\omega_p$ and (b) $293\omega_p$. The frequency for both cases is $\Omega=1.1$.}\label{fig:c}
\end{center}
\end{figure}

\begin{figure}[tbh]
\begin{center}
\includegraphics[width=80mm]{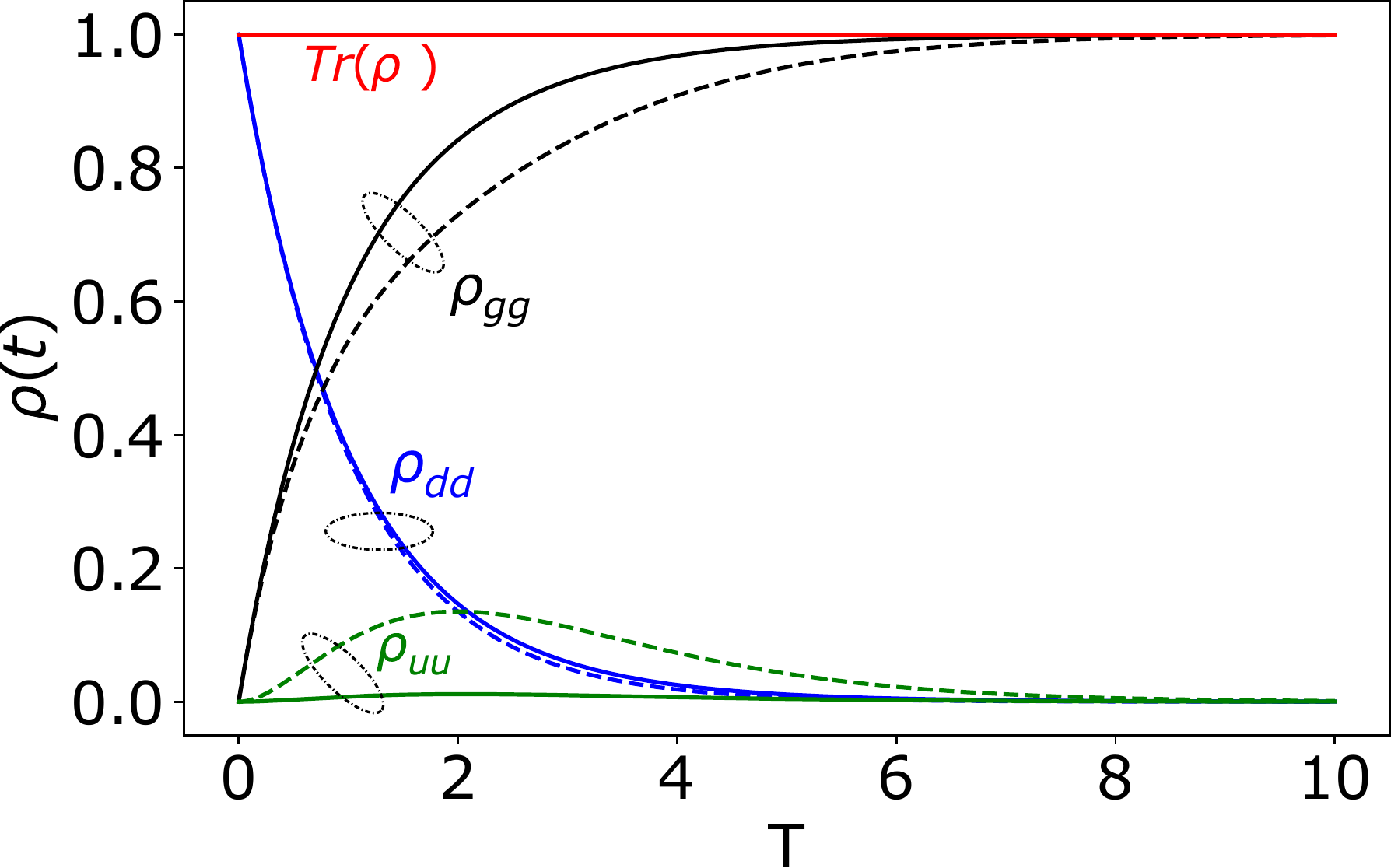}
\caption{The matrix elements of the density matrix of the qubits as a function of $T$ for the case of the Weyl semimetal (dashed lines) and the Dirac semimetal (solid lines). The bulk plasmon frequency of the normal metal is $5\omega_p$ and $\Omega=1.1$. }\label{fig:rh}
\end{center}
\end{figure}

In Fig. \ref{fig:c}, we show $\mathcal{C}$ as a function of $T\equiv 2\Gamma t/\hbar^2$ using SPP in the Weyl semimetal (dashed line) and in the Dirac semimetal (solid line) at the same $\Omega=1.1$. In the case of the Dirac semimetal, the SPP is not unidirectional as shown in Fig. \ref{fig:disp} (d) and (e). In this case, we evaluate $\Gamma_{ij}$ in Eq.~(\ref{eq:gij1}) for the reciprocal case by using polar coordinate [See Appendix \ref{app:g}]. The concurrence has the same formula as that of the nonreciprocal case, but $\rho_{ud}$ is solved numerically from Eq.~(\ref{eq:rhod}). In Fig.  \ref{fig:c} (a), we show $\mathcal{C}$ for the case of a normal metal with the bulk plasmon frequency of $5\omega_p$ and (b) $293\omega_p$ (9.6 eV). In both cases, the maximum $\mathcal{C}$ for the Weyl semimetal is of $e^{-1}=0.37$, which occurs at $T=1$. From Eq. (\ref{eq:solnres}), we understand that the maximum $\mathcal{C}$ occurs at $T=1$ with amplitude of $(|\Gamma_{21}|/\Gamma) e^{-1}$. Furthermore, from Eq. (\ref{eq:gij}), we found that $\Gamma_{21}=\Gamma~\mathrm{exp} (-iq_0(r_j-r_i))$, which gives the maximum $\mathcal{C}$ of $e^{-1}$. Similar to the previous work on the magnetized metal~\cite{gangaraj2017robust}, the reciprocal SPP gives lower $\mathcal{C}$ for the same frequency. To understand this result, we plot the matrix elements of $\rho(t)$ as a function of $T$ in Fig. \ref{fig:rh} for reciprocal (solid line) and nonreciprocal (dashed line) cases. In both cases, $\rho_{dd}$, which is the density of the initial state, decays as a function of $T$ at a similar speed. However, $\rho_{gg}$, which is the density of the relaxed state for both qubits, increases faster for the reciprocal case. It means that the excited SPP in the reciprocal case leaves the system faster compared with that in the nonreciprocal case. In the case of nonreciprocal case, the excited SPP from qubit-1 can only be directed toward the qubit-2, which increases the probability of exciting the qubit-2. The probability that the qubit-2 is excited, is represented by $\rho_{uu}$. As we can see, $\rho_{uu}$ in the nonreciprocal case is much higher than that in the reciprocal case, which means that the coupling between qubits is also more significant in the nonreciprocal case giving better $\mathcal{C}$.  It is noted that in both cases, the Trace of $\rho(t)$ is always equal to unity (red lines).

\begin{figure}[tbh]
\begin{center}
\includegraphics[width=85mm]{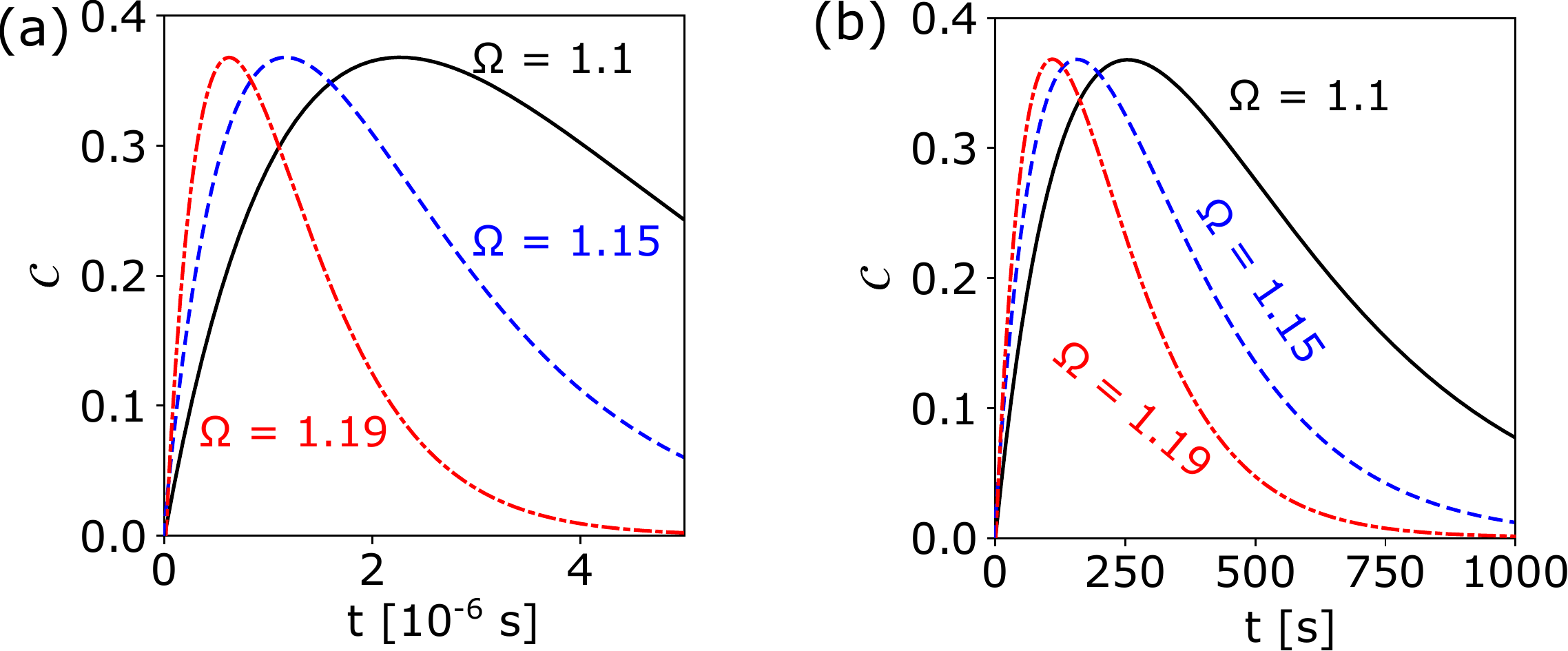}
\caption{The concurrence for the Weyl semimetal for several values of $\Omega$ as a function of $t$. (a) The bulk plasmon frequency of the normal metal is $5\omega_p$ and (b) $293\omega_p$. }\label{fig:ct}
\end{center}
\end{figure}

In Fig. \ref{fig:ct}, we show $\mathcal{C}$ for the case of the Weyl semimetal as a function of time $t$ for several values of $\Omega$. In (a), we use a normal metal with the bulk plasmon frequency of $5\omega_p$. The peak of $\mathcal{C}$ moves toward smaller $t$ with increasing $\Omega$ since $\Gamma_{ij}$ increases. Even though the maximum entanglement is reached faster for larger $\Omega$, the entanglement decays faster compared with that for smaller $\Omega$ as shown by Eq.~(\ref{eq:gij}). It is noted since the frequency of SPP in the Weyl semimetal is within the THz region, $\Gamma_{ij}$ is much smaller than that for SPP in the conventional metal, which lies at a much higher frequency (visible region). Thus, we expect a longer lifetime of the entanglement. The lifetime of the entanglement is understood as the time it takes before $\mathcal{C}$ asymptotically reaches zero~\cite{Orszag2010}. For instance, in Fig. \ref{fig:ct} (a), the lifetime of the entanglement is within $10^{-6}$s compared with that of $10^{-8}$s for magnetized metal~\cite{gangaraj2017robust}. When we keep $\Omega$, the lifetime can be increased by increasing the bulk plasmon frequency of the normal metal. In Fig. \ref{fig:ct} (b), we show the case for normal metal with bulk plasmon frequency of $293\omega_p$ (9.6 eV). The lifetime of the entanglement is much longer than in (a). The increase in the lifetime is caused by the decrease of $\alpha$. The square of $\alpha$, which is given by Eq. (\ref{eq:u}), has a unit of inverse length that determines the volume for the quantization of the SPP energy. In particular, this length corresponds to the effective length of SPP in the perpendicular direction to the surface~\cite{archambault2010quantum,gonzalez2013theory}. When we increase the dielectric constant of the normal metal, the intensity of the electric field inside the normal metal becomes much reduced due to the screening. However, since the energy should be normalized to the energy of the harmonic oscillator for a fixed number of SPP $n$, which depends only on the frequency $(\hbar\omega_q(n+1/2))$, the intensity of the electric field inside the Weyl semimetal would be increased to compensate the large screening in the normal metal. The increase in the electric field intensity inside the Weyl semimetal would increase the effective length of SPP. As a result, $\alpha$ would decrease and $\Gamma_{ij}$ becomes smaller, leading to a longer lifetime of the entanglement.
\begin{figure}[t]
\begin{center}
\includegraphics[width=85mm]{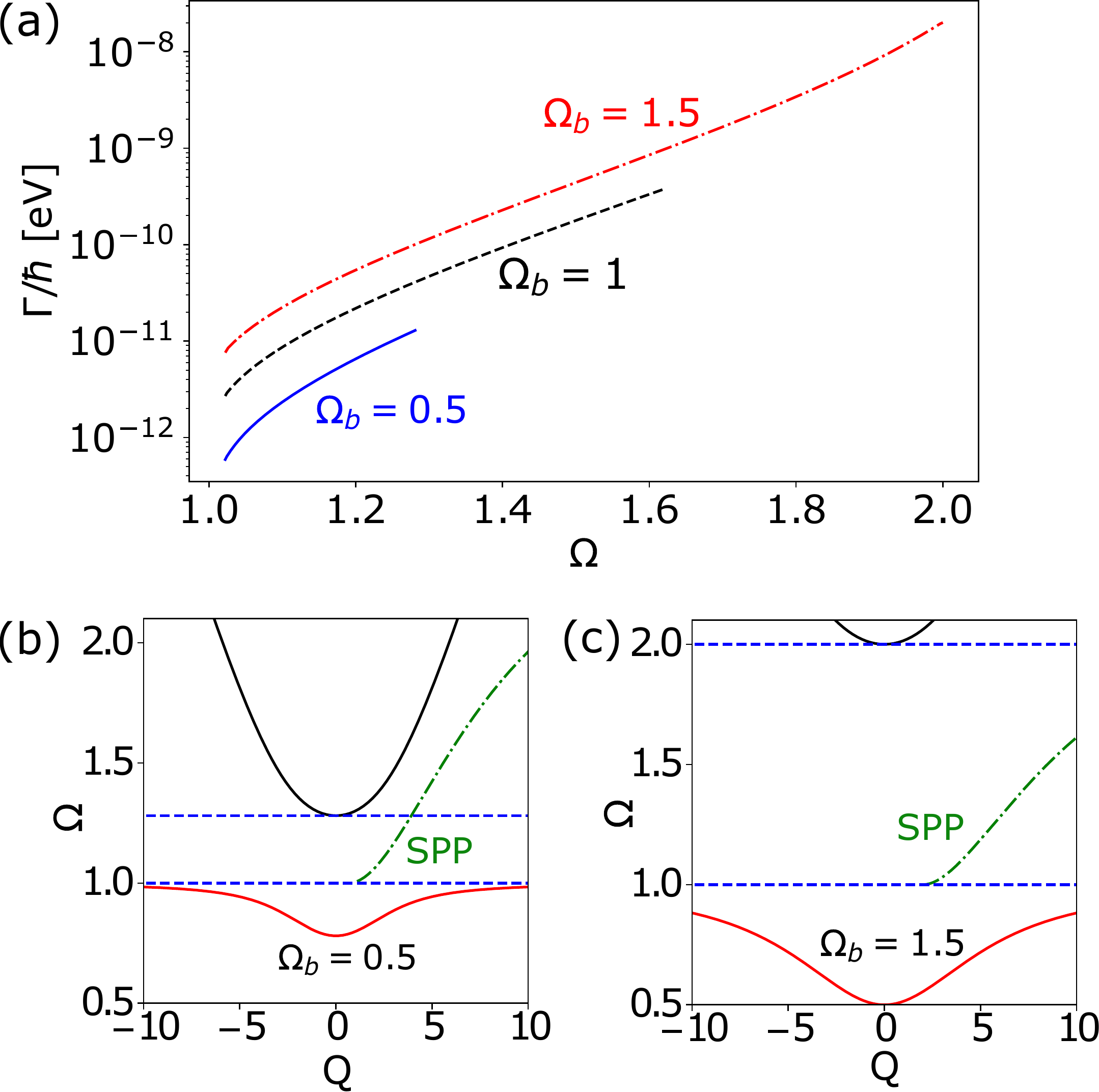}
\caption{(a) $\Gamma/\hbar$ given in eV as a function $\Omega$ for several $\Omega_b$ values. In this plot, we use a normal metal with the bulk plasmon frequency of $10\omega_p$. The frequency of TM bulk photon propagating inside the Weyl semimetal and the frequency of the SPP for (b) $\Omega_b=0.5$ and (c) $\Omega_b=1.5$. }\label{fig:gm}
\end{center}
\end{figure}

In Fig. \ref{fig:gm} (a), we show $\Gamma/\hbar$ given in unit of eV as a function of $\Omega$ for several $\Omega_b$. As shown in Eq. (\ref{eq:gap}), the band gap increases with increasing $\Omega_b$. In Fig. \ref{fig:gm} (a), the range of $\Omega$ corresponds to the band gap for the case of $\Omega_b=1.5$. Since we consider only the SPP inside the common band gap, then there are discontinuities for $\Gamma/\hbar$ for $\Omega_b<1.5$. As we can see, $\Gamma/\hbar$ increases with increasing $\Omega_b$ for a fixed $\Omega$, which means that the lifetime of entanglement becomes shorter, but the peak of entanglement is reached faster. The increase of $\Gamma$ originates from the decrease of the group velocity of the SPP when $\Omega_b$ is increased as shown by the dispersion of SPP for $\Omega_b=0.5$ and $1.5$ given in Figs. \ref{fig:gm} (b) and (c), respectively [see Eq. (\ref{eq:f})]. The decrease in the group velocity leads to the increase in the density of states of the SPP, leading to the enhancement of the qubit coupling with SPP.  

\section{Conclusions}
We have evaluated the entanglement of two qubits mediated by the nonreciprocal SPP on the surface of the Weyl semimetal. The SPP becomes nonreciprocal even without the presence of applied magnetic fields.  The nonreciprocity originates from the intrinsic Hall conductivity of the Weyl semimetal. By solving the master equation for the density matrix elements of the qubits, we found that the nonreciprocal SPP gives better entanglement  than the reciprocal one found in the Dirac semimetal. Furthermore, we can utilize the  SPP in the Weyl semimetal to control the qubits whose transition frequency is within the THz range. Although the system takes a longer time to reach maximum entanglement compared with the case of conventional metals, the entanglement here also has a longer lifetime. For a fixed frequency and bulk plasmon frequency of the Weyl semimetal, the lifetime of the entanglement can be changed by modifying the permittivity of the adjacent material or changing the separation of the Weyl nodes. 

\begin{acknowledgements}
We acknowledge Mahameru BRIN for the high-performance computing facilities. EHH acknowledges support from the National Research Fund Luxembourg under grants CORE C20/MS/14764976/TOPREL and C21/MS/15752388/NAVSQM.
\end{acknowledgements}

\appendix
\section{Formulation of $\Gamma_{ij}$ for reciprocal SPPs}
\label{app:g}
The $\Gamma_{ij}$ parameter for reciprocal SPPs can be derived similarly to that for the unidirectional SPPs given in Eq. (\ref{eq:gij1}). In this case, we use the polar coordinate for $q$ to obtain the $\Gamma_{ij}$.

\begin{align}
    \Gamma_{ij}&=\sum\limits_{k} g_{ki}g^*_{kj} \left[\pi\delta(\omega_0-\omega_k)+\frac{i}{\omega_0-\omega_k}\right]\label{eq:gijr1}\\
    &=\frac{L_x}{2\pi}\frac{L_y}{2\pi}\int\limits_{0}^{\infty} kdk\int\limits_{0}^{2\pi}d\phi g_{ki}g^*_{kj} \left[\pi\delta(\omega_0-\omega_k)+\frac{i}{\omega_0-\omega_k}\right]\\
    &=\frac{\hbar}{8\pi^2\varepsilon_0}\int\limits_{0}^{\infty} kdk\int\limits_{0}^{2\pi}d\phi \omega_k |\mathbf{u_k}(z)\cdot \mathbf{d_{12}}|^2e^{-ik\cos\phi(r_j-r_i)}\nonumber\\&\times\left[\left(\frac{\partial\omega_k}{\partial k}\right)_{k=k_0}^{-1}\pi\delta(k_0-k)+\frac{i}{\omega_0-\omega_k}\right]
    \label{eq:gaminr}\\
    &=\frac{\hbar}{8\pi^2\varepsilon_0}\int\limits_{0}^{\infty} dk~G(k,\omega_k)\nonumber\\&\times\left[\left(\frac{\partial\omega_k}{\partial k}\right)_{k=k_0}^{-1}\pi\delta(k_0-k)+\frac{i}{\omega_0-\omega_k}\right]\label{eq:gamin2},
    \end{align}
where $G(k,\omega_k)$ is defined as,
\begin{align}
    G(k,\omega_k)\equiv k\omega_k \int\limits_{0}^{2\pi}d\phi  |\mathbf{u_k}(z)\cdot \mathbf{d_{12}}|^2e^{-ik\cos\phi(r_j-r_i)}.
\end{align}
The integration of the first term inside the bracket in Eq. (\ref{eq:gamin2}) can be done easily by substituting $\omega_k$ with $\omega_0$ and $k$ with $k_0$, which is the wavevector of the SPP with energy $\omega_0$. Thus, we obtain,
\begin{widetext}
\begin{align}
    \frac{\hbar}{8\pi^2\varepsilon_0}\int\limits_{0}^{\infty} dk~G(k,\omega_k)\left(\frac{\partial\omega_k}{\partial k}\right)_{k=k_0}^{-1}\pi\delta(k_0-k)=\frac{\hbar}{8\pi\varepsilon_0}H(k_0,\omega_0)\label{eq:gijr1f},
\end{align}
where,
\begin{align}
   H(k,\omega_k)\equiv  \left(\frac{\partial\omega_k}{\partial k}\right)_{k=k_0}^{-1} G(k,\omega_k).
\end{align}
\end{widetext}
The integration of the second term inside the bracket in Eq. (\ref{eq:gamin2}) is done by using the contour integral with contour along an infinitesimal half-circle around $\omega_0$. 
\begin{widetext}
\begin{align}
    \frac{\hbar}{8\pi^2\varepsilon_0}\int\limits_{0}^{\infty} dk~\frac{iG(k,\omega_k)}{\omega_0-\omega_k}&=\frac{\hbar}{16\pi^2\varepsilon_0}\int\limits_{-\infty}^{\infty} d\omega_k~\left(\frac{\partial\omega_k}{\partial k}\right)^{-1}\frac{iG(k,\omega_k)}{\omega_0-\omega_k}\nonumber\\
    &=\frac{\hbar}{16\pi^2\varepsilon_0}\left(\frac{\partial\omega_k}{\partial k}\right)_{k=k_0}^{-1}iG(k,\omega_k)\times -i\pi=\frac{\hbar}{16\pi\varepsilon_0}H(k_0,\omega_0)
    \label{eq:gijr2}.
\end{align}
\end{widetext}
Thus, the $\Gamma_{ij}$ for the reciprocal case is given by the summation of Eqs. (\ref{eq:gijr1f}) and (\ref{eq:gijr2}).

\section{The calculated Chern number}
\label{app:c}
The expressions of electric fields are given as follows:
\begin{align}
    E_x &= \frac{1}{\omega\varepsilon_0}\frac{1}{\varepsilon_1^2-\varepsilon_2^2}\left(\varepsilon_1 k\cos\phi + i\varepsilon_2 k\sin\phi\right)\nonumber\\
    &\equiv \Gamma\left(\varepsilon_1 k\cos\phi + i\varepsilon_2 k\sin\phi\right)\\
    E_z &= \frac{1}{\omega\varepsilon_0}\frac{1}{\varepsilon_1^2-\varepsilon_2^2}\left(i\varepsilon_2 k\cos\phi - \varepsilon_1 k\sin\phi\right)\nonumber\\
    &\equiv \Gamma\left(i\varepsilon_2 k\cos\phi - \varepsilon_1 k\sin\phi\right),
\end{align}
where we take $H_0$ to be unity and define $\Gamma\equiv \frac{1}{\omega\varepsilon_0}\frac{1}{\varepsilon_1^2-\varepsilon_2^2}$. To find the $A_\phi$ as defined in Eq. (\ref{eq:ath}) of the manuscript, we calculate the derivative of the fields as follows:
\begin{align}
    \partial_\phi E_x=\Gamma\left(-\varepsilon_1 k\sin\phi + i\varepsilon_2 k\cos\phi\right)\\
        \partial_\phi E_z=\Gamma\left(-i\varepsilon_2 k\sin\phi - \varepsilon_1 k\cos\phi\right).
\end{align}
Since the $\partial_\phi H$ vanishes, only the electric fields determine the $A_\phi$. The derivative of material matrix for the dielectric function can be expressed as follows:
\begin{align}
    \partial_\omega\omega\bar{\varepsilon}=\varepsilon_0\begin{bmatrix}
\partial_\omega\omega\varepsilon_1 & 0 & i\partial_\omega\omega\varepsilon_2\\
0&\partial_\omega\omega\varepsilon_1 &  0\\
-i\partial_\omega\omega\varepsilon_2 & 0 & \partial_\omega\omega\varepsilon_1
\end{bmatrix}=\varepsilon_0\begin{bmatrix}
\Tilde{\varepsilon}_1 & 0 & 0\\
0&\Tilde{\varepsilon}_1 &  0\\
0 & 0 & \Tilde{\varepsilon}_1
\end{bmatrix}
.
\label{eq:eps}
\end{align}
It is noted that $\omega\varepsilon_2$ is constant in $\omega$ since the for WSM $\varepsilon_2\propto 1/\omega$. Thus, the derivative of material matrix for the dielectric function is diagonal. Then, we have
\begin{align}
    \mathrm{Re}(i\mathbf{E^*}\cdot\partial_\omega(\omega\bar{\varepsilon}(\omega))\frac{1}{k}\partial_\phi \mathbf{E}=-2\Gamma^2\varepsilon_0\Tilde{\varepsilon}_1k\varepsilon_1\varepsilon_2.
\end{align}
The normalization can be calculated easily and is given by
\begin{align}
    \mathbf{E^*}\cdot\partial_\omega(\omega\bar{\varepsilon}(\omega))\mathbf{E} + \mathbf{H^*}\cdot\mu_0 \mathbf{H}=\Gamma^2\varepsilon_0\Tilde{\varepsilon}_1 k^2(\varepsilon^2_1+\varepsilon^2_2)+\mu_0
.\end{align}
Thus, the $A_\phi k$ is given by
\begin{align}
   A_\phi k=-\frac{2\Gamma^2\varepsilon_0\Tilde{\varepsilon}_1\varepsilon_1\varepsilon_2k^2}{\Gamma^2 k^2\varepsilon_0\Tilde{\varepsilon}_1(\varepsilon^2_1+\varepsilon^2_2)+\mu_0}\approx-\frac{2\varepsilon_1\varepsilon_2}{(\varepsilon^2_1+\varepsilon^2_2)}.
\end{align}

At a large wavevector, for the upper branch, the frequency approaches infinity and the $\varepsilon_2\rightarrow 0$ giving vanishing $A_\phi k$. For the lower branch. the frequency approaches $\omega_p$ at a large wavevector, which makes $\varepsilon_1=0$, which also gives vanishing $A_\phi k$. Therefore, the Chern number is determined only by the $A_\phi k$ at $k\rightarrow 0$. 

The Chern number is expressed as
\begin{align}
    C_n=\lim_{k\rightarrow 0}\frac{2\varepsilon_1\varepsilon_2}{(\varepsilon^2_1+\varepsilon^2_2)}.
\end{align}
$C_n$ is an integer when $\varepsilon_1=\pm\varepsilon_2$, which occurs at the following frequencies:
\begin{align}
    \Omega_{\pm}=\frac{1}{2}(\pm\Omega_B+\sqrt{\Omega^2_B+4}).
\end{align}
In fact, the $\Omega_{\pm}$ are nothing but the frequencies of the upper ($+$) and lower ($-$) branches at $k=0$, respectively. In the case of $\Omega_B=0.5$ as is used in the manuscript, the $\Omega_+=1.28$ and $\Omega_-=0.78$. Therefore, the Chern number is an integer for both branches. The $C_n=-1$ for lower branch since $\varepsilon_1<0$ at $k=0$ and $C_n=+1$ for lower branch since $\varepsilon_1>0$ at $k=0$.
%

\end{document}